\documentclass[12pt,a4paper]{article}
\usepackage[T1,T2A]{fontenc}
\usepackage[utf8]{inputenc}
\usepackage[russian]{babel}
\usepackage{indentfirst}
\usepackage[a4paper,top=2cm,bottom=2cm,left=3cm,right=2cm,marginparwidth=1.75cm]{geometry}
\usepackage{titlesec}
\titlelabel{\thetitle.\,\,}
\usepackage{amsmath}
\usepackage{amssymb}
\usepackage{graphicx}
\usepackage{color}
\usepackage{enumitem}
\usepackage{tabularx}

\title{\textbf{СИНХРОНИЗАЦИЯ И УПРАВЛЕНИЕ ДВИЖЕНИЕМ АНСАМБЛЯ МОБИЛЬНЫХ АГЕНТОВ}}
\author{Е.М. Варварин, Г.В. Осипов}
\date{Ноябрь 2023}

\begin{document}

\maketitle

\setlength{\parindent}{2.5cm} \setlength{\hangindent}{2cm} В данной статье предлагаются способы реализации последовательного, параллельного и в виде заданной конфигурации движения ансамбля (роя) мобильных агентов  с использованием эффекта хаотической фазовой синхронизации. Показывается возможность управления движением ансамбля и определяются условия устойчивости полученных структур. \\

\setlength{\parindent}{2.5cm} \setlength{\hangindent}{2cm} \textbf{Ключевые слова}: мобильный агент, синхронизация, управление движением ансамбля, осциллятор Рёсслера.

\section{Введение}

\setlength{\parindent}{0.6cm}
Использование ансамбля мобильных агентов для изучения и анализа коллективной динамики в последние годы широко применяется в различных областях науки и техники \cite{1}. Главным объектом большинства исследований в области коллективной динамики является синхронизация \cite{2}, которая сильно зависит от топологии связей ансамбля \cite{3}. Можно выделить три основные типа связей в ансамблях: локальная (связь с ближайшими соседями), нелокальная (связь не только с ближайшими соседями), глобальная (связь по принципу   “каждый с каждым” \cite{4}). Чаще всего такие связи имеют стационарный характер, т.е. топология  и сила связей не меняются во времени.  Однако в реальном мире топология большинства структур не является постоянной, связи между элементами могут появляться и исчезать, усиливаться или ослабевать \cite{5}. Системы, в которых помимо силы связи может изменяться и положение узлов, удобно рассматривать как ансамбли мобильных агентов \cite{6}. Таким образом удалось рассмотреть: синхронизацию мобильных роботов \cite{7}, локализацию объектов распределённой следящей системой \cite{8} и другие работы. В работе \cite{9} представлены результаты управления поведением ансамбля мобильных агентов на плоскости.

\section{Модель}

В качестве мобильного агента рассмотрим точку, движущуюся в трёхмерном пространстве $(x,y,z)$  так, что её траектория полностью совпадает с траекторией соответствующего ей хаотического осциллятора. В данной работе, не теряя общности, рассмотрим осциллятор Рёсслера (1).

\begin{equation}
    \begin{cases}
    \dot{x}_i = - w_i y_i - z_i\\
    \dot{y}_i = w_i x_i + a_i y_i,\quad i=\overline{1,N}\\
    \dot{z}_i = b_i + z_i (x_i - c_i)\\
    \end{cases}
\end{equation}
где $a_i, b_i, c_i$ -- положительные параметры. В последующих экспериментах примем $a_i = 0.22, b_i = 0.1, c_i = 8.5$, параметр, характеризующие временные масштабы осцилляций -- $w_i \in [0.93; 1.07]$.

Организацию управления движением ансамбля мобильных агентов в пространстве можно разбить на два этапа: установление определённой конфигурации агентов и выведение агентов на заданную траекторию движения.

Для решения поставленных задач мы используем  методы: хаотической фазовой синхронизации (для задания ансамблю агентов определённой конфигурации их расположения в трёхмерном пространстве) и вынужденной синхронизацией (для обеспечения достижения заданного закона движения в пространстве). Под хаотической фазовой синхронизацией понимается процесс установления при достаточно сильной связи между взаимодействующими хаотическими осцилляторами одинаковой усреднённой частоты колебаний и ограниченной по модулю разности фаз (фазовому сдвигу) осцилляторов \cite{10}. 

Целью нашего исследования является создание такого ансамбля агентов, в котором  взаимодействия агентов с соседями начинаются только при их достаточной, наперёд заданной  близости, поэтому при любой конфигурации связь между агентами $i$ - тым и $j$ - тым будет удовлетворять условию (3):

\begin{equation}
    d = 
    \begin{cases}
        d', \: (x_i - x_j)^2 + (y_i - y_j)^2 + (z_i - z_j)^2 < r^2 \\
        0, \: \text{в противном случае}
    \end{cases}
\end{equation}
где $d'=const$ -- параметр связи. В нашем исследовании примем: $d' = 0.2, r=4$. Т.е. агенты начинают взаимодействовать при попадании в шар радиуса $r$.

\section{Реализованные конфигурации движения роя мобильных агентов.}
\subsection{Последовательное движение агентов.}
    Добавим в систему (1) «притягивающую» связь по координате $y$ следующим образом:
    
    \begin{equation}
        \begin{cases}
        \dot{x}_i = - w_i y_i - z_i\\
        \dot{y}_i = w_i x_i + a_i y_i + \sum_j d(y_j-y_i),\quad i=\overline{1,N}\\
        \dot{z}_i = b_i + z_i (x_i - c_i)\\
        \end{cases}
    \end{equation}
    
    Начальные условия всех осцилляторов разные. С течением времени в силу хаотичности движений изображающие точки рано или поздно сближаются на расстояние меньше $r$. Между агентами возникает взаимодействие и, если связь достаточно сильная, то агенты синхронизуются. Далее все большее число агентов сближаются на расстояние $r$ и синхронизуются. Образуются кластеры синхронизованных агентов. Фазовые траектории при этом близки, но имеет место фазовый сдвиг. В итоге достигается глобальная хаотическая фазовая синхронизация,  в результате  которой агенты двигаются  друг за другом «по цепочке».  Результат численных экспериментов (глобальная  синхронизация агентов) представлен на Рис. 1а.

\subsection{Параллельное движение агентов.}
    Кроме последовательно движения агентов удаётся реализовать их параллельное движение – движение в виде «единого строя». Параллельное движение достигается с помощью добавления «отталкивающей» связи по координате $x$:

    \begin{equation}
        \begin{cases}
        \dot{x}_i = - w_i y_i - z_i + \sum_j d(x_i-x_j) + \sum_j \frac{d}{(x_j - x_i)}\\
        \dot{y}_i = w_i x_i + a_i y_i + \sum_j d(y_j-y_i),\qquad \qquad \qquad i=\overline{1,N}\\
        \dot{z}_i = b_i + z_i (x_i - c_i)\\
        \end{cases}
    \end{equation}
    
    Аналогично случаю последовательного движения с течением времени образуются пары, тройки и т.д. параллельно двигающихся агентов. В отличие от последовательно движения  при сближении $i$-го и $j$-го агентов введённая описанным выше способом связь приводит к появлению противоположно направленные силы взаимодействия, что заставляет агентов находится на определённом перпендикулярном к движению расстоянии. Т.е. за счёт «отталкивающей» связи агенты начинают двигаться одним рядом, параллельно друг другу. Поведение элементов при данной связи проиллюстрировано на Рис. 1б.
    
\subsection{Придание рою мобильных агентов структуры различных геометрических форм.}
    В данном разделе мы используем  определённые комбинации связей последовательного и параллельного движения для получения структур различной геометрической формы (прямоугольник, круг, треугольник и др.) Не теряя общности, рассмотрим конфигурацию типа «прямоугольник».\\
    Для организации подобного движения введём параметры $k_{col}$ и $k_{str}$ — число элементов в одной строке и одном столбце соответственно. Тогда для каждой отдельной строки нам нужно добавить «отталкивающую» связь. В то же время нам нужно связать строку с соседними строками «притягивающей» связью:

    \begin{equation}
        \begin{cases}
        \dot{x}_i = - w_i y_i - z_i + \sum\limits_{j=s(i)}^{s(i)+k_{str}-1} \Big[ d(x_j-x_i) + \frac{d}{(x_i - x_j)} \Big] = f_{grid}\\
        \dot{y}_i = w_i x_i + a_i y_i + \sum\limits_{j=max(0; s(i)-k_{str})}^{min(N;s(i)+2 \cdot k_{str} - 1)}d(y_j-y_i) = g_{grid},\qquad \quad i=\overline{1,N}\\
        \dot{z}_i = b_i + z_i (x_i - c_i) = h_{grid}\\
        \end{cases}
    \end{equation}

    где $k_{str} \cdot k_{col} = N, s(i) = (i-i \, mod(k_{str})+1$ -- первый элемент текущей строки для элемента $i$, $i \, mod(k_{str})$ -- остаток от деления $i$ на $k_{str}$.
    С течением времени за счёт синхронизации агенты, связанные как «притягивающей», так и «отталкивающей» связями начинают формировать группы последовательно и параллельно двигающихся агентов синхронизируются в группы. Далее происходит  объединение агентов  в единый  кластер в виде структуры, приведённой на Рис. 1в.

\section{Анализ влияния нарушения межэлементных связей.}
В данном разделе мы рассматриваем результаты по надёжности существования заданных геометрических структур ансамблей агентов. Т.е. определяются зависимости структурной устойчивости ансамбля при удалении из него части агентов. Рассмотрим два способа выбивания агентов из структуры. 
\begin{enumerate}[label=\alph*)]
    \item Выбивание элементов из центра роя.\\
    Путём численных экспериментов было обнаружено, что при любой структуре роя существует критическое значение числа удалённых агентов, при котором структура разбивается на несколько кластеров. В результате экспериментов на структуре 10х10 при удалении 12 элементов  уже наступает разделение структуры на два кластера (Таблица 1). Интересно отметить, что при определённых условиях появляются уединённые агенты, которое за все время наблюдения не примыкают ни к одному из кластеров. 
    \item Выбивание случайных элементов.\\
    В данном эксперименте элементы удаляются из случайных позиций. Теперь количество объединённых групп агентов зависит не только от числа удалённых элементов, но и от их позиций. Не трудно увидеть из системы (5), что при удалении целой строки агентов глобальная связь ансамбля теряется и рой разобьётся как минимум на две части - до и после удалённой строки. Результаты представлены в Таблица 2.
\end{enumerate}

\section{Управление роем.}
Рассмотрим задачу деактивации всех агентов - перемещение их в заданную точку пространства. Для этого помимо агентов, заданных системой (1), вводится ещё  один агент, который движется по заданной траектории. В качестве такого агента возьмём осциллятор Ван-дер-Поля:
\begin{equation}
    \begin{cases}
        \dot X = -Y\\
        \dot Y = WX + \mu(1- X^2)Y
    \end{cases}
\end{equation}

где $\mu$ - отрицательный параметр, $W = 1$. При заданном наборе параметров система (6) имеет единственный аттрактор - устойчивое состояние равновесия в точке (0,0). Именно в эту точку должен прийти весь ансамбль. Чем больше (по модулю) $\mu$, тем быстрее ансамбль попадает в заданную точку. Очевидно, что управляющая траектория может быть любой. Это может быть заданное регулярное или хаотическое движение.\\
Для остальных агентов возьмём уравнение (5) и добавим всем элементам связь с уравнением Ван-дер-Поля следующим образом:

\begin{equation}
    \begin{cases}
        \dot{x}_i = f_{grid}\\
        \dot{y}_i = g_{grid} + D(Y-y_j),\quad i=\overline{1,N}\\
        \dot{z}_i = h_{grid}\\
    \end{cases}
\end{equation}

где связь $D$ работает аналогично связи $d$, но при сближении мобильного агента с агентом, движущимся согласно уравнению Ван-дер-Поля. В результате все агенты приходят в окрестность состояния равновесия (0,0)  (Рис. 2).

\section{Результаты}

В результате исследования синхронизации и управления коллективной динамикой роя мобильных хаотических агентов -  осциллятора Рёсслера оказалось возможным получить заданные типы движений мобильных агентов в трёхмерном пространстве: последовательное – один за другим на определённом расстоянии, которым можно управлять, параллельное – «единым фронтом» и движение в виде заданных геометрических структур. Все предложенные способы структурообразования роя мобильных агентов можно рассматривать как процессы самоорганизации. Было рассмотрено влияние количества выбиваемые агентов из ансамбля (за счёт разрыва межэлементных связей) на полученную структуру. Также продемонстрировано, что при помощи «внешнего» агента возможно задать рою требуемую траекторию движения. Например, «посадить» рой в заданной точке. 

\section{Благодарности}

Работа выполнена при финансовой поддержке гранта РНФ \#23-12-00180 (задача синхронизации) и проекта № 0729-2020-0036 Министерства науки и высшего образования Российской Федерации (задача управления)

\newpage

\begin{figure}[h]
    \includegraphics[height=5.1cm]{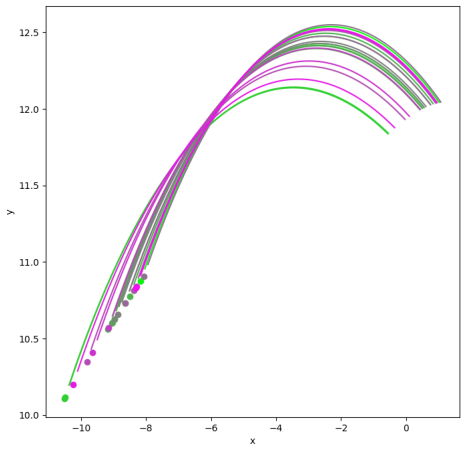}
    \includegraphics[height=5.1cm]{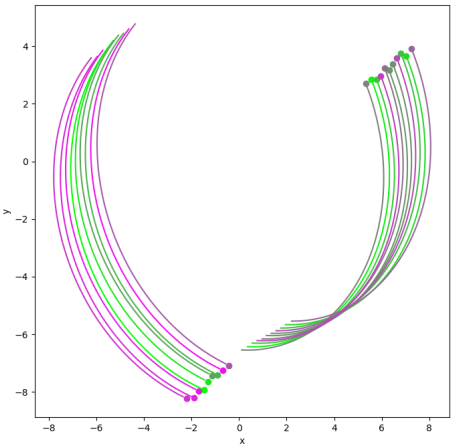}
    \includegraphics[height=5.1cm]{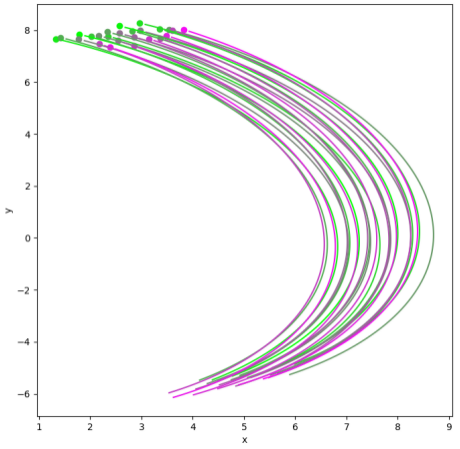}
    \centering
    \caption{\label{Рис. }  Синхронизация ансамбля мобильных агентов. Слева -- полная синхронизация при реализации последовательного движения. В центре -- кластерная синхронизация роя при параллельном движении (в дальнейшем два кластера объединятся в один). Справа -- придание рою структуры квадрата 5х5 элементов.}
\end{figure}

\begin{figure}[h]
    \includegraphics[height=5cm]{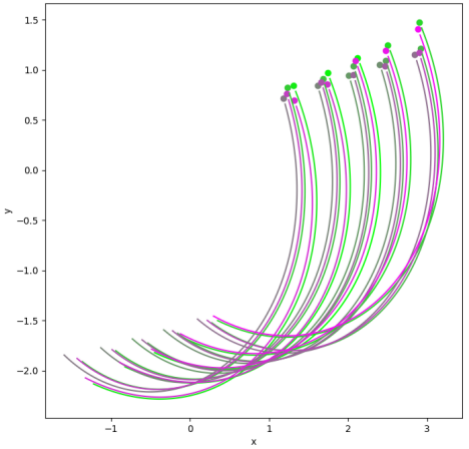}
    \includegraphics[height=5cm]{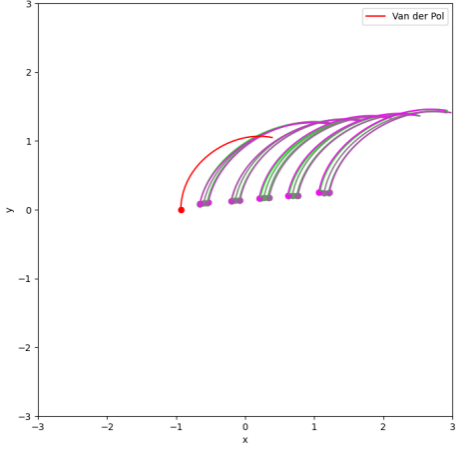}
    \includegraphics[height=5cm]{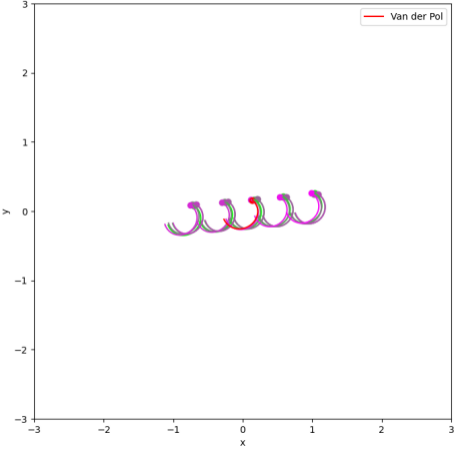}
    \centering
    \caption{\label{Рис. } Захват ансамбля агентом, заданным уравнением Ван-дер-Поля. Слева -- состояние роя до добавления нового агента. В центре -- новый элемент начал притягивать всех агентов. Справа -- рой движется по траектории фокуса к состоянию равновесия агента, заданного уравнением Ван-дер-Поля. Значение параметра $\mu = -0.2$.}
\end{figure}

\newpage

\begin{table}[]
    \caption{ Зависимость числа устойчивых кластеров от числа удалённых элементов из ансамбля, имеющего структуру квадрата 10x10 элементов. Все элементы удалялись из центра структуры.}
    \begin{tabular} {|p{3cm}|c|c|c|c|c|c|c|c|p{2cm}|p{2cm}|r|}
        \hline
        Кол-во удалённых элементов & 4 & 8 & 12 & 16 & 24 & 32 & 36 & 44 & 52 & 60 & 64 \\
        \hline
        Кол-во сохранившихся кластеров & 1 & 1 & 2 & 2 & 2 & 3 & 3 & 2 & 2 и 2 уединённых & 3 и 1 уединённый & 4 \\
        \hline
    \end{tabular}\\
\end{table}

\begin{table}[]
\caption{Зависимость числа устойчивых кластеров  от числа удалённых элементов из ансамбля, имеющего структуру квадрата 10x10 элементов. Элементы удалялись случайно.}
    \begin{tabular}{|l|l|l|l|l|l|l|l|l|l|l|l|}
        \hline
        Кол-во удалённых элементов     & 10 & 15 & 20 & 25 & 30 & 35 & 40 & 45 & 50 & 55 & 60 \\ \hline
        Кол-во сохранившихся кластеров & 1 & 1 & 1 & 1 & 1 & 1 & 1 & 2 & 2 & 2 & 2  
        \\ \hline
    \end{tabular}
\end{table}
\end{document}